# Toward molecular neuroeconomics of obesity.

Taiki Takahashi[1]

[1] Department of Behavioral Science, Hokkaido University

Corresponding Author: Taiki Takahashi

Email: taikitakahashi@gmail.com

Department of Behavioral Science, Hokkaido University

N.10, W.7, Kita-ku, Sapporo, 060-0810, Japan

TEL: +81-11-706-3057   FAX: +81-11-706-3066

**Acknowledgements:** The research reported in this paper was supported by a grant from the Grant- in-Aid for Scientific Research ("global center of excellence" grant) from the Ministry of Education, Culture, Sports, Science and Technology of Japan.

Summary:

Because obesity is a risk factor for many serious illnesses such as diabetes, better understandings of obesity and eating disorders have been attracting attention in neurobiology, psychiatry, and neuroeconomics. This paper presents future study directions by unifying (i) economic theory of addiction and obesity (Becker and Murphy, 1988; Levy 2002; Dragone 2009), and (ii) recent empirical findings in neuroeconomics and neurobiology of obesity and addiction. It is suggested that neurobiological substrates such as adiponectin, dopamine (D2 receptors), endocannabinoids, ghrelin, leptin, nesfatin-1, norepinephrine, orexin, oxytocin, serotonin, vasopressin, CCK, GLP-1, MCH, PYY, and stress hormones (e.g., CRF) in the brain (e.g., OFC, VTA, NAcc, and the hypothalamus) may determine parameters in the economic theory of obesity. Also, the importance of introducing time-inconsistent and gain/loss-asymmetrical temporal discounting (intertemporal choice) models based on Tsallis' statistics and incorporating time-perception parameters into the neuroeconomic theory is emphasized. Future directions in the application of the theory to studies in neuroeconomics and neuropsychiatry of obesity at the molecular level, which may help medical/psychopharmacological treatments of obesity (e.g., with sibutramine), are discussed.

1. **Introduction:**

There is a trend of a substantial rise in obesity in the world population [1]. Problematically, the relationship between sweetened drinks, obesity, and diabetes has also been established [2]. Moreover, some investigators have recently emphasized the similarities between obesity and drug addiction [3]. In the field of economics, in order to model addictive behavior, Becker and Murphy proposed an economic theory of addiction by incorporating temporal discounting processes [4], and subsequent theoretical work further developed economic theories of obesity and eating disorders, to demonstrate the importance of temporal discounting processes in obesity [5,6]. Because (i) economic theory of addiction has extensively and successfully been utilized in behavioral neuroeconomics of addiction [7,8] and (ii) theoretical frameworks of temporal discounting plays a pivotal role in investigations into obesity and binge eating [9,10,11,12], in addition to harmful addiction [7], it might be a promising direction to employ economic theory of obesity and eating disorders, which incorporates temporal discounting into decision-making processes, for a better understanding of neurobiological bases of obesity. In this study, I present how neuroeconomic theory of obesity will help elucidate the neurobiological bases of obesity at the molecular level (i.e., "molecular neuroeconomics" of obesity), and how recently developed temporal discounting model based on Tsallis' statistics will be utilized to study problematic eating behavior due to irrationality in temporal discounting [12]. Furthermore, for a better understanding of obesity at the molecular level, the roles of neurobiological substrates (e.g., adiponectin, leptin, ghrelin, dopamine, serotonin, norepinephrine, orexin, melanocyte-concentrating hormone (MCH) and stress hormones) in the brain regions (e.g., the ventral tegmental area, the nucleus accumbens, and the hypothalamus) in determining the parameters in the economic theory of obesity are suggested.

This paper is organized in the following manner. In Section 2, I briefly introduce economic theory of addiction and a recently-developed temporal discounting model based on Tsallis' statistics. Moreover, findings in neurobiology regarding the molecular mechanisms of obesity and addiction are briefly reviewed. In Section 3, I proposed a unified framework for a molecular neuroeconomic theory of obesity, and explain how neurobiological substrates may modulate parameters in the neuroeconomic theory of obesity. In Section 4, some conclusions from this study and future study directions by utilizing the present molecular neuroeconomic theory, and how to test the present theory experimentally in future neuroeconomic studies are discussed.

2. **Economic theory of obesity**

**2.1 Economic theory of rational obesity**

Following Becker and Murphy's economic theory of addiction [4], Levy (2002) [5] proposed an economic theory of rational obesity. The model assumes "rational" decision-making processes in that an agent is supposed to be time-consistent (i.e., exponential discounting) and forward-looking. In Levy's model, the agent is supposed to maximize

$$J = \int_0^T e^{-\rho t} U(c(t)) \Phi (W(t) - W^*)^2 dt, \tag{1}$$

where $\Phi$ indicates the probability of living beyond life expectancy $t \in [0, T]$ (T is a temporal horizon), $\rho$ is an exponential time-discount rate (i.e., consistent time-discounting: $e^{-\rho t}$ is assumed), U is a utility from consumption c(t) (food intake), W is a body weight, and W* is a physiologically-optimal body weight. Note that deviations from the physiologically-optimal body weight shorten lifespan. Levy [5] further put simple assumptions for W, U, and $\Phi$:

$$W'(t) = c(t) - \delta W(t), \tag{2}$$

where $\delta$ is a "depreciation rate" of body weight which indicates the effect of body weight on burning calories, for the utility function U, it is assumed that

$$U(c(t)) = c(t)^\beta, \tag{3}$$

where $\beta \in (0,1)$ is the utility elasticity ("saliency of food intake"), and for $\Phi$, it is assumed that,

$$\Phi = \Phi_0 \exp[-\mu (W-W^*)^2], \tag{4}$$

where $\mu$ is a positive real parameter (the rate of decline of probability of living beyond t due to deviations of body weight from W) and $\Phi_0 \in (0,1)$. Note that parameter $\mu$ corresponds to "the cost of obesity" in terms of lifespan. By solving the optimal problem with these assumptions (equations 2, 3, and 4), Levy [5] showed that the steady-state body weight $W_{ss}$ is:

$$W_{ss} = 0.5W^* + 0.5[W^{*2} + \frac{2\beta(\rho+\delta)}{\delta\mu}]^{0.5}. \tag{5}$$

This indicates that rational body weight $W_{ss}$ is larger than physiologically-optimal body weight $W^*$ ($W_{ss} > W^*$), which may explain an increased prevalence in obesity in the world as foods have become more readily available. Furthermore, equation 5 implies that subjects with steeper temporal discounting (i.e., larger $\rho$) may be more obese. Consistent with the economic theory of obesity, some recent studies in behavioral economics have revealed that obesity is associated with steeper temporal discounting [9,10,11,12]. However, to date, little is known how neurobiological substrates control parameters in equation 5, although the theory has been extendedly analyzed [6]. It is also to be noted that the economic theory of obesity differs from the economic theory of addiction in that Levy's economic theory of obesity does not incorporate habit-formation processes, which has a pivotal role in the economic theory of addiction [4]. Namely, food is assumed to be non-addictive in Levy's economic theory of obesity. A more recent model by Dragone (2009) [6] incorporates habit-forming (addictive) food intake.

## 2.2 Irrationality in temporal discounting

In the economic theory of obesity introduced above, temporal discounting is assumed to be rational (time-consistent), which is reflected in the exponential discounting ($e^{-\rho t}$) in equation 1. However, accumulating empirical evidence suggests that human temporal discounting behavior is irrational (time-inconsistent) [8]. In order to describe impulsivity and irrationality in temporal discounting, the q-exponential time-discount model for delayed rewards has been introduced and experimentally examined [13,14]:

$$V_{q+}(D) = V_{q+}(0) / \exp_{q+}(k_{q+}D) = V_{q+}(0) / [1+(1-q+)k_{q+}D]^{1/(1-q+)} \tag{6}$$

where $V_{q+}(D)$ is the subjective value of a reward obtained at delay D, $q+$ is a parameter indicating irrationality in temporal discounting for gain (smaller $q+<1$ values correspond to more irrational discounting for delayed gains), and $k_{q+}$ is a parameter of impulsivity regarding the reward at delay $D=0$ (q-exponential discount rate at delay $D=0$). Note that when $q+=0$, equation 6 is the same as a hyperbolic discount function, while $q+\rightarrow 1$, is the same as an exponential discount function in equation 1. Furthermore, it is known that delayed gains and losses are distinctly processed in the brain [15] and loss is less steeply temporally-discounted than gains, which is referred to as the "sign effect" [12]. Therefore, we should prepare the q-exponential discount

function for delayed loss:

$$V_{q-}(D) = V_{q-}(0) / \exp_{q-}(k_{q-}D) = V_{q-}(0)/[1+(1-q-)k_{q-}D]^{1/(1-q-)} \tag{7}$$

where $V_{q-}(D) > 0$ is the (absolute, unsigned) subjective value (subjective magnitude) of a loss at delay D, $q-$ is a parameter indicating irrationality in temporal discounting for loss (smaller q-<1 values correspond to more irrational discounting for delayed losses), and $k_{q-}$ is a parameter of impulsivity regarding the loss (i.e., degree of procrastination) at delay $D=0$. A recent behavioral health economic study demonstrated that obesity is associated with both the sign effect and time-inconsistency, in addition to impulsivity (time-discount rate) [12]. This indicates the necessity of the utilization of the q-exponential models for gain and loss, for the optimization problem in the economic theory of obesity.

## 2.3 Neurobiological substrates of obesity and addiction

Recent studies implied that there are common neurobiological pathways for obesity and addiction. The sensitivity or reactivity of the 'common reward pathway' is affected by several biological factors such as the density of dopamine receptors, the amount of dopamine released into the synapse, and the rapidity of its transport back into the cell by the re-uptake protein. Individual differences in reward sensitivity have been strongly implicated in the risk for drug addictions as well as compulsive overeating [3].

In addition to dopamine, several neurobiological substrates (e.g., leptin and ghrelin) are shown to be related to obesity. While studies of drug addiction have focused on mesolimbic and mesocortical dopamine circuits, research on obesity and food intake has largely focused on the hypothalamus. In studies on obesity, analysis of hypothalamic mechanisms was accelerated by the identification of key molecules, such as leptin and the leptin receptor, that control body weight. Leptin (a key molecule controlling food intake in the hypothalamus) administration alters intracranial self-stimulation of brain reward systems and leptin receptor activation in the ventral tegmental area (VTA) reduces dopamine neuronal firing and food intake, suggesting an interaction with dopamine circuits [16]. Also, D2 receptors in the striatum may be involved in prefrontal metabolism [17]. This indicates the importance of employing a reward value-based neuroecononomic theory for investigations into obesity.

Regarding the control of metabolic processes, adiponectin controls glucose regulation and fatty acid catabolism. Adiponectin also modulates cannabinoid receptors in the hypothalamic nuclei which mediate energy balance and body weight, as well as in the mesolimbic system which mediates the incentive value of food [18]. The lateral hypothalamic area synthesizes orexin/hypocretin, which control food intake, wakefulness, and energy expenditure. Furthermore, orexin/hypocretin and corticotropin releasing factor (CRF) also modulate drug-related dopamine activities [19], again indicating links between obesity and addiction.

In the hypothalamus, melanin-concentrating hormone (MCH) receptors also distribute. MCH is implicated in the regulation of different physiological functions, including energy homeostasis and mood that is supported by the distribution of MCH and MCHR1 in the hypothalamus as well as corticolimbic structures. Chung et al., (2009) [20] recently demonstrated that MCH is also involved in cocaine addiction. Therefore, MCH should also be considered in developing neuroeconomic theory of obesity in relation to addiction [4,6].

Concerning brain regions involved in food intake, activation of orbitofrontal cortex (OFC), medial frontal cortex (MFC), amygdala, hippocampal formation, and insula are consistently reported to be associated with hunger and incentive value of food [21]. A recent neuroeconomic study (Hare et al., 2009) also reported that ventromedial prefrontal cortex valuation system is activated during food consumption decision-making.

In the field of neuroeconomics, the roles of oxytocin (a type of nona-neuropeptides) in social preferences have been attracting attention [23,24,25,26,27]. Oxytocin is known to be related to empathy and social stress coping [28,29,30]. Interestingly, oxytocin has recently been indicated to regulate obesity as well. For instance, it was reported that oxytocin receptor-deficiency is associated with obesity [31,32]. Also, nesfatin-1 (an anorexigenic peptide) neurons co-express with oxytocin, vasopressin and MCH in the hypothalamus, indicating important roles of oxytocin and its interactions with other hormones in the regulation of food intake [33].

With respect to neuroendocrine regulation of food intake, cholecystokinin (CCK), glucagon-like peptide-1 (GLP-1) and peptide YY (PYY), act as satiety signals [34]. Ghrelin also targets the hypothalamus to regulate food intake and adiposity. Ghrelin acts on the VTA and modulate neuronal functions, food intake and reward processing [35], indicating that neuroeconomic theories of obesity should consider the roles of ghrelin in dopamine systems.

3. **Toward a molecular neuroeconomic theory of obesity**
In this section, I propose frameworks for neuroeconomic theory of obesity and addiction. First of all, as introduced in 2.3, neurobiological substrates related to energy metabolism

(e.g., adiponectin), which may determine $\delta$ in equation 5 are strongly related to fat, food intake and body weight [36], consistent with the prediction from the economic theory of obesity (see equation 5, which indicates $\delta$ controls the deviation of body weight from the physiologically-optimal weight). Also, hypothalamic hormones may additionally modulate $\beta$ (the shape of the utility function) which determines the incentive value of food, possibly by altering dopaminergic activities via the activation of endocannabinoid receptors. Furthermore, because both obesity and addiction are associated with steeper temporal discounting, it is important to examine how hypothalamic hormones and adiponectine modulate $\rho$ (a time preference parameter) in equation 5. Note that larger $\rho$ (steeper temporal discounting) is predicted to associate with obesity from equation 5. Also, Charlton et al., (2008) [37] suspected metabolic rate may be related to time-discount rate. Therefore, molecular neuroeconomic studies should examine how molecular neurobiological substrates modulating energy metabolism control time-discount rate. This investigation may help to develop molecular neuroeconomics of intertemporal choice incorporating the roles of metabolic molecules. Overall, future studies should investigate how adiponectine, stress hormones, and hypothalamic hormones modulate parameters in equation 5 via various receptors in the hypothalamus, and dopaminergic regions such as VTA and NAcc.

With respect to the relationship between irrationality in decision over time and obesity, it is important to utilize temporal discounting models which incorporate time-inconsistency and gain/loss asymmetry. One possible direction is to utilize the q-exponential discount models for gain and loss (equations 6 and 7), for the optimization problem in equation 1. The model for optimization is:

$$J_q = \int_0^T [V_{q+}(t) - V_{q-}(t)] \Phi(W(t) - W^*)^2 dt, \qquad (8)$$

where $V_{q+}(t)$ and $V_{q-}(t)$ are q-exponential discount functions for gain and loss (equation 6 and 7), respectively. This model may significantly change the characteristics of optimized body weight, because the roles of time-inconsistency and the sign effect are incorporated. Therefore, future neuroeconomic studies should examine how hypothalamic hormones, in addiction to adiponectine and stress hormones control parameters in the q-exponential discounting. Moreover, a serotonin-norepinephrine reuptake inhibitor sibutramine is known to be effective in medical treatments for obesity. Because our previous studies indicate that serotonin and norepinephrine are strongly associated with parameters in the temporal discounting function [38,39], it is possible that sibutramine reduces body weight through a decrease in time-discount rates and time-inconsistency which are mediated by serotonin and norepinephrine.

As stated earlier, brain regions such as OFC and MFC are associated with incentive value of food. It is therefore possible that the activations of these brain regions are correlated with $\beta$ in equation 3 and 5. Furthermore, as stated earlier, peptides CCK, GLP-1 and PYY act as satiation signals. Therefore, future neuroeconomic studies should investigate how these peptides control $\beta$ in equation 3.

Regarding neuropsychological processing underlying irrationality in temporal discounting, it has been proposed that nonlinearity in temporal cognition may explain time-inconsistent intertemporal choice [40,41,42], which has later been confirmed experimentally [43]. Because obesity is related to time-inconsistency in temporal discounting [12], it is expected that (i) more obese subjects have more impaired time-perception and (ii) hunger is associated with impaired temporal cognition, which have already been reported [44,45]. Also, some nutrients modify the speed of internal clock [46]. Hence, future neuroeconomic studies on obesity should examine how food intake-controlling hormones modulate temporal cognition.

Finally, as shown in equation 5, "the cost of obesity" in terms of lifespan (parametrized with $\mu$) may reduce body weight. Hence, future behavioral economic studies should test whether a reduction in the cost of obesity (e.g., mortality due to diabetes and coronary heart diseases) due to progress in medical treatment for obesity-related diseases actually increases obesity or not.

4. **Implications for neuroeconomics and neurobiolgy of obesity and addiction**

This study is the first to present a possible unified framework for molecular neuroeconomic theory of obesity. Furthermore, possible future theoretical and experimental study directions were proposed. For a better understanding of the relationship between obesity and addiction, neurobiological studies utilizing transgenic animals with modifications in the receptors in the hypothalamus and dopaminergic regions may be helpful. Future studies could conduct a food intake task, a temporal cognition and discounting tasks, and an experiment of self-administration of addictive drugs in the same transgenic animals. After identifying which neurobiological substrates most strongly and distinctly control parameters in the neuroeconomic model of obesity (e.g., equation 8), future clinical treatments can be focused on psychopharmacological medications which normalize the parameters.